\documentclass[conference]{IEEEtran}
\IEEEoverridecommandlockouts
\usepackage{cite}
\usepackage{amsmath,amssymb,amsfonts}
\usepackage{algorithmic}
\usepackage{graphicx}
\usepackage{subcaption} 
\usepackage{textcomp}
\usepackage{xcolor}
\usepackage{tabularx} 
\usepackage{makecell} 
\usepackage{pdflscape}
\usepackage{array}
\usepackage{enumitem}

\usepackage[acronyms,nonumberlist,nopostdot,nomain,nogroupskip,acronymlists={hidden}]{glossaries}
\newacronym{3gpp}{3GPP}{3rd Generation Partnership Project}
\newacronym{4g}{4G}{4th generation}
\newacronym{5g}{5G}{5th generation}
\newacronym{6g}{6G}{6th generation}
\newacronym{5gc}{5GC}{5G Core}
\newacronym{adc}{ADC}{Analog to Digital Converter}
\newacronym{aerpaw}{AERPAW}{Aerial Experimentation and Research Platform for Advanced Wireless}
\newacronym{ai}{AI}{Artificial Intelligence}
\newacronym{aimd}{AIMD}{Additive Increase Multiplicative Decrease}
\newacronym{am}{AM}{Acknowledged Mode}
\newacronym{amc}{AMC}{Adaptive Modulation and Coding}
\newacronym{amf}{AMF}{Access and Mobility Management Function}
\newacronym{aops}{AOPS}{Adaptive Order Prediction Scheduling}
\newacronym{api}{API}{Application Programming Interface}
\newacronym{apn}{APN}{Access Point Name}
\newacronym{ap}{AP}{Application Protocol}
\newacronym{aqm}{AQM}{Active Queue Management}
\newacronym{ausf}{AUSF}{Authentication Server Function}
\newacronym{avc}{AVC}{Advanced Video Coding}
\newacronym{awgn}{AGWN}{Additive White Gaussian Noise}
\newacronym{balia}{BALIA}{Balanced Link Adaptation Algorithm}
\newacronym{bbu}{BBU}{Base Band Unit}
\newacronym{bdp}{BDP}{Bandwidth-Delay Product}
\newacronym{ber}{BER}{Bit Error Rate}
\newacronym{bf}{BF}{Beamforming}
\newacronym{bler}{BLER}{Block Error Rate}
\newacronym{brr}{BRR}{Bayesian Ridge Regressor}
\newacronym{bs}{BS}{Base Station}
\newacronym{bsr}{BSR}{Buffer Status Report}
\newacronym{bss}{BSS}{Business Support System}
\newacronym{ca}{CA}{Carrier Aggregation}
\newacronym{caas}{CaaS}{Connectivity-as-a-Service}
\newacronym{cav}{CAV}{Connected and Autonoums Vehicle}
\newacronym{cb}{CB}{Code Block}
\newacronym{cc}{CC}{Congestion Control}
\newacronym{ccid}{CCID}{Congestion Control ID}
\newacronym{cco}{CC}{Carrier Component}
\newacronym{cd}{CD}{Continuous Delivery}
\newacronym{cdd}{CDD}{Cyclic Delay Diversity}
\newacronym{cdf}{CDF}{Cumulative Distribution Function}
\newacronym{cdn}{CDN}{Content Distribution Network}
\newacronym{cli}{CLI}{Command-line Interface}
\newacronym{cn}{CN}{Core Network}
\newacronym{codel}{CoDel}{Controlled Delay Management}
\newacronym{comac}{COMAC}{Converged Multi-Access and Core}
\newacronym{cord}{CORD}{Central Office Re-architected as a Datacenter}
\newacronym{cornet}{CORNET}{COgnitive Radio NETwork}
\newacronym{cosmos}{COSMOS}{Cloud Enhanced Open Software Defined Mobile Wireless Testbed for City-Scale Deployment}
\newacronym{cots}{COTS}{Commercial Off-the-Shelf}
\newacronym{cp}{CP}{Control Plane}
\newacronym{cyp}{CP}{Cyclic Prefix}
\newacronym{up}{UP}{User Plane}
\newacronym{cpu}{CPU}{Central Processing Unit}
\newacronym{cqi}{CQI}{Channel Quality Information}
\newacronym{cr}{CR}{Cognitive Radio}
\newacronym{cran}{CRAN}{Cloud \gls{ran}}
\newacronym{crs}{CRS}{Cell Reference Signal}
\newacronym{csi}{CSI}{Channel State Information}
\newacronym{csirs}{CSI-RS}{Channel State Information - Reference Signal}
\newacronym{cu}{CU}{Central Unit}
\newacronym{d2tcp}{D$^2$TCP}{Deadline-aware Data center TCP}
\newacronym{d3}{D$^3$}{Deadline-Driven Delivery}
\newacronym{dac}{DAC}{Digital to Analog Converter}
\newacronym{dag}{DAG}{Directed Acyclic Graph}
\newacronym{das}{DAS}{Distributed Antenna System}
\newacronym{dash}{DASH}{Dynamic Adaptive Streaming over HTTP}
\newacronym{dc}{DC}{Dual Connectivity}
\newacronym{dccp}{DCCP}{Datagram Congestion Control Protocol}
\newacronym{dce}{DCE}{Direct Code Execution}
\newacronym{dci}{DCI}{Downlink Control Information}
\newacronym{dctcp}{DCTCP}{Data Center TCP}
\newacronym{dl}{DL}{Downlink}
\newacronym{dmr}{DMR}{Deadline Miss Ratio}
\newacronym{dmrs}{DMRS}{DeModulation Reference Signal}
\newacronym{drlcc}{DRL-CC}{Deep Reinforcement Learning Congestion Control}
\newacronym{dsrc}{DSRC}
{dedicated short-range communications}
\newacronym{d2d}{D2D}{device-to-device}
\newacronym{drs}{DRS}{Discovery Reference Signal}
\newacronym{du}{DU}{Distributed Unit}
\newacronym{e2e}{E2E}{end-to-end}
\newacronym{earfcn}{EARFCN}{E-UTRA Absolute Radio Frequency Channel Number}
\newacronym{ecaas}{ECaaS}{Edge-Cloud-as-a-Service}
\newacronym{ecn}{ECN}{Explicit Congestion Notification}
\newacronym{edf}{EDF}{Earliest Deadline First}
\newacronym{embb}{eMBB}{Enhanced Mobile Broadband}
\newacronym{empower}{EMPOWER}{EMpowering transatlantic PlatfOrms for advanced WirEless Research}
\newacronym{enb}{eNB}{evolved Node Base}
\newacronym{endc}{EN-DC}{E-UTRAN-\gls{nr} \gls{dc}}
\newacronym{epc}{EPC}{Evolved Packet Core}
\newacronym{eps}{EPS}{Evolved Packet System}
\newacronym{es}{ES}{Edge Server}
\newacronym{etsi}{ETSI}{European Telecommunications Standards Institute}
\newacronym[firstplural=Estimated Times of Arrival (ETAs)]{eta}{ETA}{Estimated Time of Arrival}
\newacronym{eutran}{E-UTRAN}{Evolved Universal Terrestrial Access Network}
\newacronym{faas}{FaaS}{Function-as-a-Service}
\newacronym{fapi}{FAPI}{Functional Application Platform Interface}
\newacronym{fdd}{FDD}{Frequency Division Duplexing}
\newacronym{fdm}{FDM}{Frequency Division Multiplexing}
\newacronym{fdma}{FDMA}{Frequency Division Multiple Access}
\newacronym{fed4fire}{FED4FIRE+}{Federation 4 Future Internet Research and Experimentation Plus}
\newacronym{fir}{FIR}{Finite Impulse Response}
\newacronym{fit}{FIT}{Future \acrlong{iot}}
\newacronym{fpga}{FPGA}{Field Programmable Gate Array}
\newacronym{fr2}{FR2}{Frequency Range 2}
\newacronym{fr1}{FR1}{Frequency Range 1}
\newacronym{fs}{FS}{Fast Switching}
\newacronym{fscc}{FSCC}{Flow Sharing Congestion Control}
\newacronym{ftp}{FTP}{File Transfer Protocol}
\newacronym{fw}{FW}{Flow Window}
\newacronym{ge}{GE}{Gaussian Elimination}
\newacronym{gnb}{gNB}{Next Generation Node Base}
\newacronym{gop}{GOP}{Group of Pictures}
\newacronym{gpr}{GPR}{Gaussian Process Regressor}
\newacronym{gpu}{GPU}{Graphics Processing Unit}
\newacronym{gtp}{GTP}{GPRS Tunneling Protocol}
\newacronym{gtpc}{GTP-C}{GPRS Tunnelling Protocol Control Plane}
\newacronym{gtpu}{GTP-U}{GPRS Tunnelling Protocol User Plane}
\newacronym{gtpv2c}{GTPv2-C}{\gls{gtp} v2 - Control}
\newacronym{gw}{GW}{Gateway}
\newacronym{harq}{HARQ}{Hybrid Automatic Repeat reQuest}
\newacronym{hetnet}{HetNet}{Heterogeneous Network}
\newacronym{hh}{HH}{Hard Handover}
\newacronym{hol}{HOL}{Head-of-Line}
\newacronym{hqf}{HQF}{Highest-quality-first}
\newacronym{hss}{HSS}{Home Subscription Server}
\newacronym{http}{HTTP}{HyperText Transfer Protocol}
\newacronym{ia}{IA}{Initial Access}
\newacronym{iab}{IAB}{Integrated Access and Backhaul}
\newacronym{ic}{IC}{Incident Command}
\newacronym{ietf}{IETF}{Internet Engineering Task Force}
\newacronym{imsi}{IMSI}{International Mobile Subscriber Identity}
\newacronym{imt}{IMT}{International Mobile Telecommunication}
\newacronym{iot}{IoT}{Internet of Things}
\newacronym{ip}{IP}{Internet Protocol}
\newacronym{itu}{ITU}{International Telecommunication Union}
\newacronym{kpi}{KPI}{Key Performance Indicator}
\newacronym{kpm}{KPM}{Key Performance Measurement}
\newacronym{kvm}{KVM}{Kernel-based Virtual Machine}
\newacronym{los}{LoS}{Line of Sight}
\newacronym{lsm}{LSM}{Link-to-System Mapping}
\newacronym{lstm}{LSTM}{Long Short Term Memory}
\newacronym{lte}{LTE}{Long Term Evolution}
\newacronym{lxc}{LXC}{Linux Container}
\newacronym{m2m}{M2M}{Machine to Machine}
\newacronym{mac}{MAC}{Medium Access Control}
\newacronym{manet}{MANET}{Mobile Ad Hoc Network}
\newacronym{mano}{MANO}{Management and Orchestration}
\newacronym{mc}{MC}{Multi-Connectivity}
\newacronym{mcc}{MCC}{Mobile Cloud Computing}
\newacronym{mchem}{MCHEM}{Massive Channel Emulator}
\newacronym{mcs}{MCS}{Modulation and Coding Scheme}
\newacronym{mec2}{MEC}{Multi-access Edge Computing}
\newacronym{mec}{MEC}{Mobile Edge Computing}
\newacronym{mfc}{MFC}{Mobile Fog Computing}
\newacronym{mgen}{MGEN}{Multi-Generator}
\newacronym{mi}{MI}{Mutual Information}
\newacronym{mib}{MIB}{Master Information Block}
\newacronym{miesm}{MIESM}{Mutual Information Based Effective SINR}
\newacronym{mimo}{MIMO}{Multiple Input, Multiple Output}
\newacronym{ml}{ML}{Machine Learning}
\newacronym{mlr}{MLR}{Maximum-local-rate}
\newacronym[plural=\gls{mme}s,firstplural=Mobility Management Entities (MMEs)]{mme}{MME}{Mobility Management Entity}
\newacronym{mmtc}{mMTC}{Massive Machine-Type Communications}
\newacronym{mmwave}{mmWave}{millimeter wave}
\newacronym{mpdccp}{MP-DCCP}{Multipath Datagram Congestion Control Protocol}
\newacronym{mptcp}{MPTCP}{Multipath TCP}
\newacronym{mr}{MR}{Maximum Rate}
\newacronym{mrdc}{MR-DC}{Multi \gls{rat} \gls{dc}}
\newacronym{mse}{MSE}{Mean Square Error}
\newacronym{mss}{MSS}{Maximum Segment Size}
\newacronym{mt}{MT}{Mobile Terminal}
\newacronym{mtd}{MTD}{Machine-Type Device}
\newacronym{mtu}{MTU}{Maximum Transmission Unit}
\newacronym{mumimo}{MU-MIMO}{Multi-user \gls{mimo}}
\newacronym{mvno}{MVNO}{Mobile Virtual Network Operator}
\newacronym{nalu}{NALU}{Network Abstraction Layer Unit}
\newacronym{nas}{NAS}{Network Attached Storage}
\newacronym{nat}{NAT}{Network Address Translation}
\newacronym{nbiot}{NB-IoT}{Narrow Band IoT}
\newacronym{nfv}{NFV}{Network Function Virtualization}
\newacronym{nfvi}{NFVI}{Network Function Virtualization Infrastructure}
\newacronym{ni}{NI}{Network Interfaces}
\newacronym{nic}{NIC}{Network Interface Card}
\newacronym{now}{NOW}{Non Overlapping Window}
\newacronym{nsm}{NSM}{Network Service Mesh}
\newacronym{nr}{NR}{New Radio}
\newacronym{nrf}{NRF}{Network Repository Function}
\newacronym{nsa}{NSA}{Non Stand Alone}
\newacronym{nse}{NSE}{Network Slicing Engine}
\newacronym{nssf}{NSSF}{Network Slice Selection Function}
\newacronym{o2i}{O2I}{Outdoor to Indoor}
\newacronym{oai}{OAI}{OpenAirInterface}
\newacronym{oaicn}{OAI-CN}{\gls{oai} \acrlong{cn}}
\newacronym{oairan}{OAI-RAN}{\acrlong{oai} \acrlong{ran}}
\newacronym{oam}{OAM}{Operations, Administration and Maintenance}
\newacronym{ofdm}{OFDM}{Orthogonal Frequency Division Multiplexing}
\newacronym{olia}{OLIA}{Opportunistic Linked Increase Algorithm}
\newacronym{omec}{OMEC}{Open Mobile Evolved Core}
\newacronym{onap}{ONAP}{Open Network Automation Platform}
\newacronym{onf}{ONF}{Open Networking Foundation}
\newacronym{onos}{ONOS}{Open Networking Operating System}
\newacronym{oom}{OOM}{\gls{onap} Operations Manager}
\newacronym{opnfv}{OPNFV}{Open Platform for \gls{nfv}}
\newacronym{oran}{O-RAN}{Open \gls{ran}}
\newacronym{orbit}{ORBIT}{Open-Access Research Testbed for Next-Generation Wireless Networks}
\newacronym{os}{OS}{Operating System}
\newacronym{oss}{OSS}{Operations Support System}
\newacronym{pa}{PA}{Position-aware}
\newacronym{pase}{PASE}{Prioritization, Arbitration, and Self-adjusting Endpoints}
\newacronym{pawr}{PAWR}{Platforms for Advanced Wireless Research}
\newacronym{pbch}{PBCH}{Physical Broadcast Channel}
\newacronym{pcef}{PCEF}{Policy and Charging Enforcement Function}
\newacronym{pcfich}{PCFICH}{Physical Control Format Indicator Channel}
\newacronym{pcrf}{PCRF}{Policy and Charging Rules Function}
\newacronym{pdcch}{PDCCH}{Physical Downlink Control Channel}
\newacronym{pdcp}{PDCP}{Packet Data Convergence Protocol}
\newacronym{pdsch}{PDSCH}{Physical Downlink Shared Channel}
\newacronym{pdu}{PDU}{Packet Data Unit}
\newacronym{pf}{PF}{Proportional Fair}
\newacronym{pgw}{PGW}{Packet Gateway}
\newacronym{phich}{PHICH}{Physical Hybrid ARQ Indicator Channel}
\newacronym{phy}{PHY}{Physical}
\newacronym{pmch}{PMCH}{Physical Multicast Channel}
\newacronym{pmi}{PMI}{Precoding Matrix Indicators}
\newacronym{powder}{POWDER}{Platform for Open Wireless Data-driven Experimental Research}
\newacronym{ppo}{PPO}{Proximal Policy Optimization}
\newacronym{ppp}{PPP}{Poisson Point Process}
\newacronym{prach}{PRACH}{Physical Random Access Channel}
\newacronym{prb}{PRB}{Physical Resource Block}
\newacronym{psnr}{PSNR}{Peak Signal to Noise Ratio}
\newacronym{pss}{PSS}{Primary Synchronization Signal}
\newacronym{pucch}{PUCCH}{Physical Uplink Control Channel}
\newacronym{pusch}{PUSCH}{Physical Uplink Shared Channel}
\newacronym{qam}{QAM}{Quadrature Amplitude Modulation}
\newacronym{qci}{QCI}{\gls{qos} Class Identifier}
\newacronym{qoe}{QoE}{Quality of Experience}
\newacronym{qos}{QoS}{Quality of Service}
\newacronym{quic}{QUIC}{Quick UDP Internet Connections}
\newacronym{ra}{RA}{Resouces Allocation}
\newacronym{rach}{RACH}{Random Access Channel}
\newacronym{ran}{RAN}{Radio Access Network}
\newacronym[firstplural=Radio Access Technologies (RATs)]{rat}{RAT}{Radio Access Technology}
\newacronym{rbg}{RBG}{Resource Block Group}
\newacronym{rcn}{RCN}{Research Coordination Network}
\newacronym{rc}{RC}{RAN Control}
\newacronym{rec}{REC}{Radio Edge Cloud}
\newacronym{red}{RED}{Random Early Detection}
\newacronym{renew}{RENEW}{Reconfigurable Eco-system for Next-generation End-to-end Wireless}
\newacronym{rf}{RF}{Radio Frequency}
\newacronym{rfc}{RFC}{Request for Comments}
\newacronym{rfr}{RFR}{Random Forest Regressor}
\newacronym{ric}{RIC}{\gls{ran} Intelligent Controller}
\newacronym{rlc}{RLC}{Radio Link Control}
\newacronym{rlf}{RLF}{Radio Link Failure}
\newacronym{rlnc}{RLNC}{Random Linear Network Coding}
\newacronym{rmr}{RMR}{RIC Message Router}
\newacronym{rmse}{RMSE}{Root Mean Squared Error}
\newacronym{rnis}{RNIS}{Radio Network Information Service}
\newacronym{rr}{RR}{Round Robin}
\newacronym{rrc}{RRC}{Radio Resource Control}
\newacronym{rrm}{RRM}{Radio Resource Management}
\newacronym{rru}{RRU}{Remote Radio Unit}
\newacronym{rs}{RS}{Remote Server}
\newacronym{rsrp}{RSRP}{Reference Signal Received Power}
\newacronym{rsrq}{RSRQ}{Reference Signal Received Quality}
\newacronym{rss}{RSS}{Received Signal Strength}
\newacronym{rssi}{RSSI}{Received Signal Strength Indicator}
\newacronym{rtt}{RTT}{Round Trip Time}
\newacronym{ru}{RU}{Radio Unit}
\newacronym{rus}{RSU}{Road Side Unit}
\newacronym{rw}{RW}{Receive Window}
\newacronym{rx}{RX}{Receiver}
\newacronym{s1ap}{S1AP}{S1 Application Protocol}
\newacronym{sa}{SA}{standalone}
\newacronym{sack}{SACK}{Selective Acknowledgment}
\newacronym{sap}{SAP}{Service Access Point}
\newacronym{sc2}{SC2}{Spectrum Collaboration Challenge}
\newacronym{scef}{SCEF}{Service Capability Exposure Function}
\newacronym{sch}{SCH}{Secondary Cell Handover}
\newacronym{scoot}{SCOOT}{Split Cycle Offset Optimization Technique}
\newacronym{sctp}{SCTP}{Stream Control Transmission Protocol}
\newacronym{sdap}{SDAP}{Service Data Adaptation Protocol}
\newacronym{sdk}{SDK}{Software Development Kit}
\newacronym{sdm}{SDM}{Space Division Multiplexing}
\newacronym{sdma}{SDMA}{Spatial Division Multiple Access}
\newacronym{sdn}{SDN}{Software-defined Networking}
\newacronym{sdr}{SDR}{Software-defined Radio}
\newacronym{seba}{SEBA}{SDN-Enabled Broadband Access}
\newacronym{sgsn}{SGSN}{Serving GPRS Support Node}
\newacronym{sgw}{SGW}{Service Gateway}
\newacronym{si}{SI}{Study Item}
\newacronym{sib}{SIB}{Secondary Information Block}
\newacronym{sinr}{SINR}{Signal to Interference plus Noise Ratio}
\newacronym{sip}{SIP}{Session Initiation Protocol}
\newacronym{siso}{SISO}{Single Input, Single Output}
\newacronym{sla}{SLA}{Service Level Agreement}
\newacronym{sm}{SM}{Service Model}
\newacronym{smo}{SMO}{Service Management and Orchestration}
\newacronym{smsgmsc}{SMS-GMSC}{\gls{sms}-Gateway}
\newacronym{snr}{SNR}{Signal-to-Noise-Ratio}
\newacronym{son}{SON}{Self-Organizing Network}
\newacronym{sptcp}{SPTCP}{Single Path TCP}
\newacronym{srb}{SRB}{Service Radio Bearer}
\newacronym{srn}{SRN}{Standard Radio Node}
\newacronym{srs}{SRS}{Sounding Reference Signal}
\newacronym{ss}{SS}{Synchronization Signal}
\newacronym{sss}{SSS}{Secondary Synchronization Signal}
\newacronym{st}{ST}{Spanning Tree}
\newacronym{svc}{SVC}{Scalable Video Coding}
\newacronym{tb}{TB}{Transport Block}
\newacronym{tcp}{TCP}{Transmission Control Protocol}
\newacronym{tdd}{TDD}{Time Division Duplexing}
\newacronym{tdm}{TDM}{Time Division Multiplexing}
\newacronym{tdma}{TDMA}{Time Division Multiple Access}
\newacronym{tfl}{TfL}{Transport for London}
\newacronym{tfrc}{TFRC}{TCP-Friendly Rate Control}
\newacronym{tft}{TFT}{Traffic Flow Template}
\newacronym{tgen}{TGEN}{Traffic Generator}
\newacronym{tip}{TIP}{Telecom Infra Project}
\newacronym{tm}{TM}{Transparent Mode}
\newacronym{to}{TO}{Telco Operator}
\newacronym{tr}{TR}{Technical Report}
\newacronym{trp}{TRP}{Transmitter Receiver Pair}
\newacronym{ts}{TS}{Technical Specification}
\newacronym{tti}{TTI}{Transmission Time Interval}
\newacronym{ttt}{TTT}{Time-to-Trigger}
\newacronym{tx}{TX}{Transmitter}
\newacronym{uas}{UAS}{Unmanned Aerial System}
\newacronym{uav}{UAV}{Unmanned Aerial Vehicle}
\newacronym{udm}{UDM}{Unified Data Management}
\newacronym{udp}{UDP}{User Datagram Protocol}
\newacronym{udr}{UDR}{Unified Data Repository}
\newacronym{ue}{UE}{User Equipment}
\newacronym{uhd}{UHD}{\gls{usrp} Hardware Driver}
\newacronym{ul}{UL}{Uplink}
\newacronym{um}{UM}{Unacknowledged Mode}
\newacronym{uml}{UML}{Unified Modeling Language}
\newacronym{upa}{UPA}{Uniform Planar Array}
\newacronym{upf}{UPF}{User Plane Function}
\newacronym{urllc}{URLLC}{Ultra Reliable and Low Latency Communications}
\newacronym{usa}{U.S.}{United States}
\newacronym{usim}{USIM}{Universal Subscriber Identity Module}
\newacronym{usrp}{USRP}{Universal Software Radio Peripheral}
\newacronym{utc}{UTC}{Urban Traffic Control}
\newacronym{vim}{VIM}{Virtualization Infrastructure Manager}
\newacronym{vm}{VM}{Virtual Machine}
\newacronym{vnf}{VNF}{Virtual Network Function}
\newacronym{volte}{VoLTE}{Voice over \gls{lte}}
\newacronym{voltha}{VOLTHA}{Virtual OLT HArdware Abstraction}
\newacronym{vr}{VR}{Virtual Reality}
\newacronym{vran}{vRAN}{Virtualized \gls{ran}}
\newacronym{vss}{VSS}{Video Streaming Server}
\newacronym{v2x}{V2X}{vehicle-to-everything}
\newacronym{v2i}{V2I}{vehicle-to-infrastructure}
\newacronym{v2v}{V2V}{vehicle-to-vehicle}
\newacronym{v2n}{V2N}{vehicle-to-network}
\newacronym{wbf}{WBF}{Wired Bias Function}
\newacronym{wf}{WF}{Waterfilling}
\newacronym{wg}{WG}{Working Group}
\newacronym{wlan}{WLAN}{Wireless Local Area Network}
\newacronym{osm}{OSM}{Open Source \gls{nfv} Management and Orchestration}
\newacronym{pnf}{PNF}{Physical Network Function}
\newacronym{drl}{DRL}{Deep Reinforcement Learning}
\newacronym{mtc}{MTC}{Machine-type Communications}
\newacronym{osc}{OSC}{O-RAN Software Community}
\newacronym{mns}{MnS}{Management Services}
\newacronym{ves}{VES}{\gls{vnf} Event Stream}
\newacronym{ei}{EI}{Enrichment Information}
\newacronym{fh}{FH}{Fronthaul}
\newacronym{fft}{FFT}{Fast Fourier Transform}
\newacronym{laa}{LAA}{Licensed-Assisted Access}
\newacronym{plfs}{PLFS}{Physical Layer Frequency Signals}
\newacronym{ptp}{PTP}{Precision Time Protocol}
\newacronym{lidar}{LiDAR}{Light Detection And Ranging}
\newacronym{dem}{DEM}{Digital Elevation Model}
\newacronym{dtm}{DEM}{Digital Terrain Model}
\newacronym{dsm}{DEM}{Digital Surface Models}
\newacronym{ota}{OTA}{Over-The-Air}
\newacronym{ns}{NS}{Network Slicing}
\newacronym{ne}{NE}{Nash Equilibrium}
\newacronym{hf}{HF}{High Frequency}
\newacronym{noma}{NOMA}{Non-Orthogonal Multiple Access}
\newacronym{sre}{SRE}{Smart Radio Environment}
\newacronym{ris}{RIS}{Reconfigurable Intelligent Surface}
\newacronym{inp}{InP}{Infrastructure Provider}
\newacronym{smf}{SMF}{Slicing Magangement Framework}
\newacronym{nsn}{NSN}{Network Slicing Negotiation}
\newacronym{sms}{SMS}{Slicing MAC Scheduler}
\newacronym{brd}{BRD}{Best Response Dynamics}
\newacronym{dssbr}{DSSBR}{Double Step Smoothed Best Response}
\newacronym{poa}{PoA}{Price of Anarchy}
\newacronym{pos}{PoS}{Price of Stability}
\newacronym{milp}{MILP}{Mixed Integer-Linear Program}
\newacronym{pod}{PoD}{Price of DSSBR}
\newacronym{roc}{ROC}{Radio Overload Control}
\newacronym{ciot}{cIoT}{critical Internet of Things}
\newacronym{embbpr}{eMBB Pr.}{enhanced Mobile BroadBand Premium}
\newacronym{sps}{SPS}{Semi-persistent Scheduling}
\newacronym{cg}{CG}{Configured Grant}
\newacronym{embbbs}{eMBB Bs.}{enhanced Mobile BroadBand Basic}
\newacronym{en}{EN}{Edge Node}
\newacronym{ec}{EC}{Edge Computing}
\newacronym{sp}{SP}{Service Provider}
\newacronym{me}{ME}{Market Equilibrium}
\newacronym{so}{SO}{Social Optimum}
\newacronym{wso}{WSO}{Weighted Social Optimum}
\newacronym{ps}{PS}{Proportional Sharing}
\newacronym{eg}{EG}{Eisenberg-Gale program}
\newacronym{pe}{PE}{Pareto Efficiency}
\newacronym{nsw}{NSW}{Nash Social Welfare}
\newacronym{ef}{EF}{Envy-Freeness}
\newacronym{sub6}{sub-$6$GHz}{Below $6\,$GHz}
\newacronym{ncr}{NCR}{Network-Controlled Repeater}
\newacronym{nlos}{NLoS}{Non-Line of Sight}
\newacronym{src}{SRC}{Smart Radio Connection}
\newacronym{srd}{SRD}{Smart Radio Device}
\newacronym{cs}{CS}{Candidate Site}
\newacronym{tp}{TP}{Test Point}
\newacronym{fov}{FoV}{Field of View}
\newacronym{nrric}{near-RT RIC}{Near Real-time {RAN} Intelligent Controller}
\newacronym{e2ap}{E2AP}{E2 Application Protocol}
\newacronym{e2sm}{E2SM}{E2 Service Model}
\newacronym{nrtric}{non-RT RIC}{Non-Real-Time {RIC}}
\newacronym{itti}{ITTI}{Inter-task Interface}
\newacronym{bap}{BAP}{Backhaul Adaptation Protocol}
\newacronym{iabest}{IABEST}{Integrated Access and Backhaul Experimental large-Scale Tetbed}
\newacronym{teid}{TEID}{Tunnel Endpoint Identifier}
\newacronym{dlsch}{DL-SCH}{Downlink Shared Channel }
\newacronym{ulsch}{UL-SCH}{Uplink Shared Channel }
\newacronym{rsu}{RSU}{Road Side Unit}
\newacronym{its}{ITS}{Intelligent Transportation Systems}
\newacronym{vanet}{VANET}{Vehicular Ad-hoc Network}
\newacronym{dt}{DT}{Digital Twin}
\newacronym{ecc}{ECC}{Edge Computing Cluster}
\newacronym{fig}{Fig.}{Figure}
\newacronym{dnt}{DNT}{Digital Network Twin}
\newacronym{rt}{RT}{Ray Tracer}
\newacronym{cam}{CAM}{Cooperative Awareness Message}
\newacronym{prdr}{PRDR}{Packet Reception Disagreement Ratio}

\usepackage[font=small]{caption}

\begin{document}

\title{Toward Digital Network Twins:   Integrating Sionna RT in ns-3 for 6G Multi-\gls{rat} Networks Simulations}
\author{Roberto Pegurri\textsuperscript{1}, Francesco Linsalata\textsuperscript{2}, Eugenio Moro\textsuperscript{2}, Jakob Hoydis\textsuperscript{3}, Umberto Spagnolini\textsuperscript{2}\\
\small{\textit{\textsuperscript{1,2}DEIB, Politecnico di Milano, Milan, Italy}},
\small{\textit{\textsuperscript{3}{NVIDIA, France}}}\\
\small{Email: \textsuperscript{1}roberto.pegurri@mail.polimi.it, \textsuperscript{2}\{name.surname\}@polimi.it}, \textsuperscript{3}jhoydis@nvidia.com}

\maketitle

\begin{abstract}  
The increasing complexity of 6G systems demands innovative tools for network management, simulation, and optimization. This work introduces the integration of ns-3 with Sionna~RT, establishing the foundation for the first open source full-stack \gls{dnt} capable of supporting multi-\gls{rat}. By incorporating a deterministic ray tracer for precise and site-specific channel modeling, this framework addresses limitations of traditional stochastic models and enables realistic, dynamic, and multilayered wireless network simulations.
Tested in a challenging vehicular urban scenario, the proposed solution demonstrates significant improvements in accurately modeling wireless channels and their cascading effects on higher network layers. With up to 65\% observed differences in application-layer performance compared to stochastic models, this work highlights the transformative potential of ray-traced simulations for 6G research, training, and network management.
\end{abstract}


\section{Introduction}
\glsresetall
The disaggregation of 5G networks has increased the heterogeneity of components, infrastructures, and domains, significantly complicating network management—a challenge expected to grow with the advent of 6G. In response, \glspl{dnt} have emerged as promising solutions to address these issues \cite{DTMagazine}. By providing precise digital replicas of physical networks, \glspl{dnt} offer a risk-free environment for exploring innovative technologies, validating extended 6G architectures prior to deployment, and enabling real-time wireless network management in operational scenarios. These capabilities make \glspl{dnt} indispensable tools for the research and development of next-generation networks \cite{Ericsson}.

The concept of \glspl{dnt} extends beyond the simulation of networks, offering a framework to create real-time digital representations of physical systems. \glspl{dnt} integrate simulation models with live data, forming a closed-loop system where decisions about the physical entity are continuously updated based on insights from the \gls{dnt} \cite{7509384}. This paradigm becomes especially compelling in the context of 6G networks, where the complexity of the electromagnetic environment demands precise modeling of dynamic and multi-layered interactions between components \cite{10198573}. 

However, a key challenge in realizing high-fidelity \glspl{dnt} is accurately modeling the wireless propagation environment. Existing network simulators typically rely on stochastic or semi-stochastic models for channel characterization, mainly frequency-tied, which, while computationally efficient, fail to capture the nuanced physical characteristics of advanced scenarios. In contrast, ray-based propagation simulations provide a robust alternative for physical channel characterization. These methods offer detailed estimates of path loss, angles of arrival/departure, propagation delay, and Doppler shift for multipath components, making them ideal for dynamic and complex environments \cite{7152831}. Integrating these simulations into \glspl{dnt} enables the development of precise radio maps and improves their ability to reflect real world conditions, even in challenging urban or vehicular scenarios where traditional models often fall short \cite{9770941}. Furthermore, the compute architecture behind our proposal offers significant benefits in terms of interoperability and scalability. The \gls{phy} layer can be transparently accelerated using GPUs, enhancing simulation efficiency, while the disaggregation of components can exploit the economy of scale of the cloud, allowing for scalable and distributed deployments.

To maximize the impact of \gls{dnt} solutions, it is important to first identify opportunities to reuse existing functionalities, such as management functions and simulation tools, while ensuring that essential enablers, such as access to necessary data, are in place \cite{Ericsson}. Building on these developments, our work pioneers the integration of Sionna RT \cite{sionna}, an open-source \gls{rt} recognized for its computational efficiency \cite{zhu2024toward}, into the widely used ns-3 \cite{riley2010ns} simulator. This integration bridges the gap between stochastic and deterministic modeling, delivering a tool capable of supporting multi-\gls{rat}, multi-stack scenarios across different frequency bands. With its advanced channel and simulation capabilities, our approach allows for the replication of a wide range of representative real-world network deployments, producing datasets that are invaluable for training AI/ML models capable of adapting to network dynamics.

\subsection{Related works}

The computational intensity of ray-based simulations has historically limited their adoption, particularly for real-time applications \cite{zhu2024toward}. However, recent advances have addressed this bottleneck through techniques such as adaptive ray launching, simplified urban models, and efficient computational frameworks \cite{9459462, BostonTWIN}. For example, the high-fidelity Boston Digital Twin \cite{BostonTWIN} demonstrates the feasibility of integrating 3D city models into \gls{dnt}-enabled systems. 

The DeepSense 6G and DeepVerse 6G datasets \cite{Alkhateeb} advance \gls{dnt} research by combining real-world multi-modal data with high-fidelity synthetic data generated via ray tracing. However, their scenario-specific nature may limit generalizability, and their focus on the physical layer restricts applicability for higher-layer network functionalities.

In \cite{Colosseum}, the authors introduce Colosseum—a wireless network emulator with hardware-in-the-loop capabilities—as a potential \gls{dnt} platform to address key challenges and support the development of end-to-end, fully integrated, and reliable solutions. The channel emulator in Colosseum models each transceiver pair using a tapped delay line with up to four active taps. While effective for many scenarios, this approach may fall short in accurately representing environments with highly complex multipath characteristics, such as dense urban areas or cases involving intricate scattering and diffraction effects.

Implementations of \glspl{dnt} and their components are discussed in \cite{9881768}. This work primarily focuses on middleware solutions for \glspl{dnt} in aerial networks and strategies to optimize network services for replicating real-world setups. 

\subsection{Contributions}

Our main contributions are summarized as follows:

\begin{itemize} [wide]
    \item Integration of Sionna~RT into the ns-3 simulator, realizing a first-of-its-kind open-source, full-stack, and multi-\gls{rat} \gls{dnt}, overcoming the limitations of traditional stochastic and frequency-dependent channel models.
    \item Development of an adaptable and modular platform for next-generation network scenarios, featuring flexible deployment of the ray tracer with caching mechanisms and real-time synchronization, enabling efficient simulation performance and supporting the testing and optimization of 6G research. The platform’s scalability is enhanced by leveraging GPU-accelerated \gls{phy} simulations and cloud-based disaggregation for large-scale deployments.
    \item Demonstration and validation of the framework in a challenging vehicular urban scenario by comparing application-layer performance across 802.11p, LTE~\gls{v2v}, and NR~\gls{v2v} technologies, showcasing significant differences in the modeling of dynamic wireless channels. While stochastic models are well-suited for long-term performance analysis, ray tracing simulations are superior for real-time instantaneous analysis, with our results indicating a difference of up to 63\% difference in the number of received packets.
\end{itemize}

The remainder of the paper is organized as follows. Section~\ref{sec:solution} presents the proposed integration and highlights its main features. Section \ref{sec:channel} compares channel characterization in ns-3 and Sionna~RT. Numerical results demonstrating the impact of this integration at the application layer in a vehicular network are provided in Section \ref{sec:results} for both multi-\gls{rat} and multi-band scenarios. Section \ref{sec:directions} outlines potential research directions in which this work could serve as a fundamental building block. Finally, Section~\ref{sec:conclusions} concludes the paper.

\section{Integrating ns-3 with Sionna RT} \label{sec:solution}
\begin{figure}[!t]
    \centering
    \begin{subfigure}{0.5\textwidth}
        \centering
        \includegraphics[width=\linewidth]{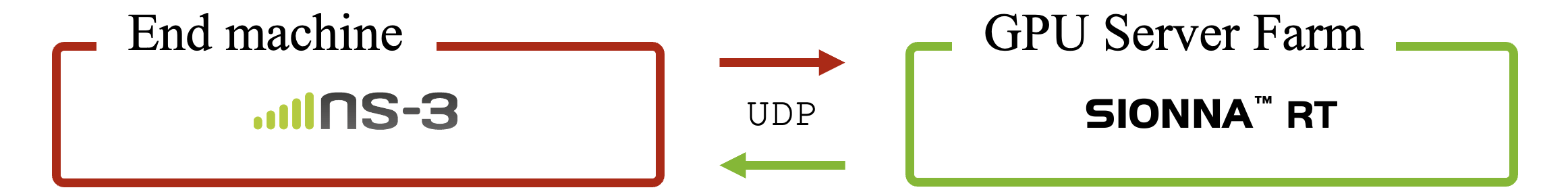}
        \caption{overview of the used setup}
        \label{setup}
    \hfill
    \end{subfigure}
    \begin{subfigure}{0.5\textwidth}
        \centering
        \includegraphics[width=\linewidth]{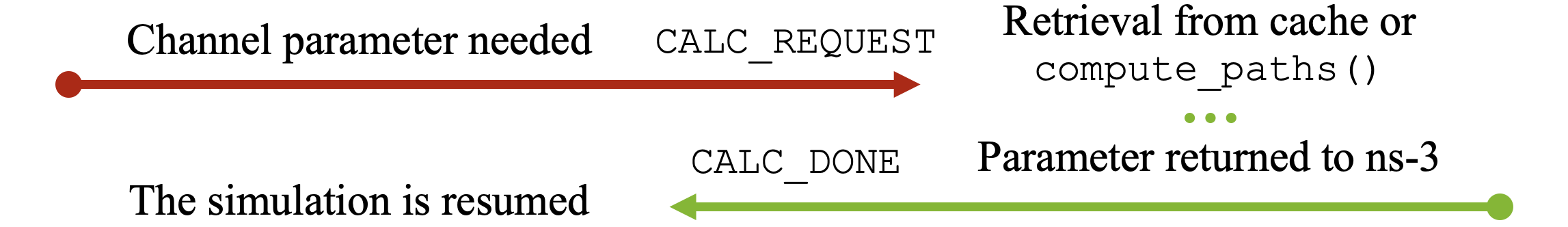}
        \caption{channel parameter exchange}
        \label{channel_parameter}
    \hfill
    \end{subfigure}
    \begin{subfigure}{0.5\textwidth}
        \centering
        \includegraphics[width=\linewidth]{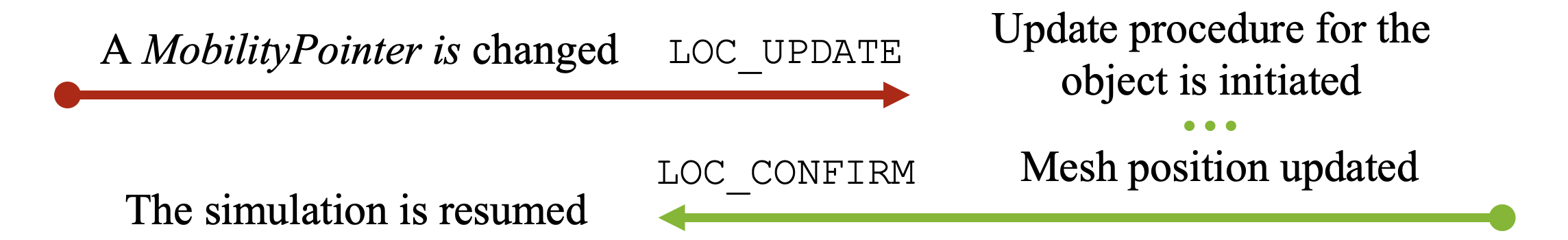}
        \caption{objects location synchronization}
        \label{location_update}
    \hfill
    \end{subfigure}
    \caption{\centering Setup overview with procedures pseudo-code.}
    \label{fig:message_exchange}
\end{figure}

While ns-3 excels in versatility and full-stack simulation, its lack of physical fidelity limits high-precision \glspl{dnt}. Integrating Sionna~RT into ns-3 as a ray tracing backbone overcomes this limitation by enabling accurate, physically consistent simulations. This section explores the integration’s modular design, communication processes, and scalability for high-precision simulations.

\subsection{Main features}
The network simulator ns-3 and Sionna~RT operate as two distinct and independent entities, communicating seamlessly via an UDP network socket.  This modular and network-based design allows the ray tracer, arguably the most computationally intensive component in a \gls{dnt}, to be deployed externally, adapting to the simulation’s specific requirements and the available infrastructure. For instance, the ray tracer can be installed on the same machine as ns-3 or, alternatively, on a dedicated GPU-accelerated server to maximize computational performance and overall scalability. 
This disaggregated approach not only provides enhanced flexibility but also opens the possibility of leveraging cost-efficient pay-as-you-go cloud services, reducing the need for upfront investments in dedicated server hardware. This adaptability makes it particularly appealing for simulations requiring large-scale computational resources without the burden of permanent infrastructure costs. For example, ns-3 could be run on a local workstation while the ray tracing computations are offloaded to a commercial cloud-based service.
To better demonstrate this capability, the setup used for the data presented in the following sections adopts a disaggregated approach, with ns-3 and Sionna~RT installed on two separate data centers (see Fig.~\ref{setup}).

\begin{table*}[!ht]
\centering
\caption{Summary of ns-3 Channel Models}
\label{tab:channel_models}
\begin{tabular}{|l|l|l|l|}
\hline
\textbf{Parameter}      & \textbf{Wi-Fi (802.11p)} & \textbf{LTE}&\textbf{NR}\\ \hline
Model Type              & Log-normal              & Semi-stochastic       &Semi-stochastic      \\ \hline
Frequency Range         & Sub-6~GHz              & Sub-6~GHz              &0.5~GHz to 100~GHz\\ \hline
\acrshort{los} Probability       & N/A                     &$\min\{\frac{18}{d},1\}\left(1-e^{-\frac{d}{36}}\right) + e^{-\frac{d}{36}}$  &3GPP TR 37.885~\cite{3gpp_tr_37_885} \\ \hline
Path Loss Formula (\acrshort{los}) & 
\begin{tabular}[c]{@{}l@{}}$Pr = Pt - Pr_0 - 10n\log_{10}(d/d_0)$\end{tabular}& 
\begin{tabular}[c]{@{}l@{}}$Pr = Pt - \text{max\{ Loss}_\text{Free Space}\text{,}$ \\ $ \text{Loss}_{LoS}(d, f_c, h_{\text{MS}}, h_{\text{BS}})$\text{\}}\end{tabular}&
\begin{tabular}[c]{@{}l@{}}$Pr = Pt - \text{Loss}_{LoS}(d, f_c)$\end{tabular}\\ \hline
Path Loss Formula (\acrshort{nlos}) & N/A &
\begin{tabular}[c]{@{}l@{}}$Pr = Pt - \text{max\{ Loss}_\text{Free Space}\text{,}$ \\ $\text{Loss}_{NLoS}(d, f_c, h_{\text{MS}}, h_{\text{BS}})$\text{\}}\end{tabular}&
\begin{tabular}[c]{@{}l@{}}$Pr = Pt - \text{Loss}_{NLoS}(d, f_c)$\end{tabular}\\ \hline
Shadowing               & None                   & None&Log-normal, $\sigma = 3 dB$\\ \hline
\end{tabular}
\end{table*}

\begin{table*}[ht!]
\centering
\caption{ns-3 path loss formulas for different scenarios in case of LTE and NR.}
\begin{tabular}{|l|l|l|}
\hline
\textbf{Scenario}     & \textbf{Path Loss Formula LTE}              & \textbf{Path Loss Formula NR} \\ \hline
\acrshort{los}& 
\begin{tabular}[c]{@{}l@{}}
$[d \leq d_{\text{BP}}]$ $22.7  \log_{10}(d) + 27 + 20  \log_{10}(f_c) $ \\ 
$[d > d_{\text{BP}}]$ $40  \log_{10}(d) + 7.56 - 17.3  \log_{10}(h_{\text{BS}} - 1) -  17.3   \log_{10}(h_{\text{MS}} - 1) + 2.7 \log_{10}(f_c) $
\end{tabular}

& $20  \log_{10}(d) + 32.4 + 20  \log_{10}(f_c)$\\ \hline

\acrshort{nlos}                  & 
\begin{tabular}[c]{@{}l@{}}
$44.9-6.55  \log_{10}(h_{\text{BS}})  \log_{10}(d) + 5.83  \log_{10}(h_{\text{BS}}) + 18.38 + 23  \log_{10}(f_c) + NLoS_{\text{Offset}}$
\end{tabular}

& $30  \log_{10}(d) + 36.85 + 18.9  \log_{10}(f_c)$\\ \hline

\end{tabular}
\label{tab:path_loss}
\end{table*}

\subsection{Operational logic}
Upon launch, the UDP socket is first established creating the communication link between the two components and enables message exchange throughout the simulation. The setup, including the parameters of the ray tracer, is fully customizable to meet specific testing requirements. For instance, it can be used to trade off latency against accuracy. 

Two primary types of communication can occur: channel parameter requests and location updates for moving objects.

\subsubsection{Channel parameter requests} These requests are initiated by ns-3 when channel details—--used to estimate received power and propagation delay—--are required, replacing the earlier reliance on internal stochastic models. For this scope, a \texttt{CALC\_REQUEST} message is created in ns-3, including the unique IDs of the involved entities (see Fig.~\ref{channel_parameter}). Consequently, the ray tracer is responsible for performing all necessary calculations with the provided information. To optimize performance, a caching mechanism for ray paths is implemented. If a valid pre-computed value exists, it is immediately retrieved and returned to ns-3 in a \texttt{CALC\_DONE} message. If the pre-comuputed value is not found, the \texttt{compute\_paths()} function in Sionna~RT calculates the required rays, which are then cached for future use. In terms of computational time, this workflow takes an average of 7.2 milliseconds using an NVIDIA A40 GPU. It also supports grouped calculations, allowing a single calculation to determine the channel parameters for all objects in the simulation. This feature is particularly advantageous to reduce GPU computation overheads.

\subsubsection{Location updates for objects} Given that the ray tracer lacks direct access to the simulation’s underlying details, any change in the positions of elements within the simulation must be promptly communicated to maintain synchronization and consistency. Location updates can originate from either mobility simulators like SUMO or from hardware-in-the-loop systems mounted on real moving objects. Each movement leads to a \texttt{MobilityPointer} update in ns-3, resulting in new positions being sent to Sionna~RT in \texttt{LOC\_UPDATE} messages, including the unique ID of the moved object (see Fig.~\ref{location_update}). If necessary, the ray tracer updates the corresponding mesh and responds with a \texttt{LOC\_CONFIRM} message. The update frequency for meshes, in terms of the minimum variation in position, can be customized. Since after a location update the scenario has changed, every pre-calculated value becomes obsolete and the ray paths cache is wiped. 

During both of these procedures the simulation is paused until the corresponding reply from the ray tracer is received.

\section{Channel Emulation Comparison} \label{sec:channel}
In this section, we detail the stochastic or semi-stochastic channel models implemented in ns-3 to simulate wireless communication environments and the main features of Sionna~RT. This is done in preparation for Sec.~\ref{sec:results}, where the network level impact of the two approaches will be evaluated.

\subsection{ns-3 Channel Modeling}
While ns-3 allows users to select any compatible channel model, our study adopts the default ns-3 models for Wi-Fi, LTE, and NR to facilitate comparisons. Table~\ref{tab:channel_models} summarizes the key parameters of the aforementioned models. Note that, in case of stochastic channel models, stock ns-3 averages the received power over several realizations.

\subsubsection{Wi-Fi}
The IEEE 802.11 family, including 802.11p, uses a log-normal propagation model with received power as:
\begin{equation}
Pr = P_t - Pr_0 - 10 n \log\left(\frac{d}{d_0}\right)
\end{equation}
where \(Pr_0\) is the power at reference distance \(d_0 = 1~\mathrm{m}\) computed using Friis equation with \(f_c = 5.89~\mathrm{GHz}\), and \(n = 3\) is the path loss exponent. This model, suited for sub-6~GHz bands, lacks \gls{nlos} considerations. 

\subsubsection{LTE and NR}
Both LTE and NR utilize semi-stochastic models based on 3GPP specifications \cite{3gpp_tr_36_885,3gpp_tr_37_885}. In both cases, the \gls{los}/\gls{nlos} conditionis first determined according to a stochastic function that depends on distance. Then, appropriate power loss functions are computed, according to the details reported in Tab.~\ref{tab:channel_models}. Notably, LTE's loss formula includes a breakpoint distance \(d_{\text{BP}} = \frac{4  (h_{\text{BS}} - 1)  (h_{\text{MS}} - 1)  f_c}{c}\) in the \gls{los} loss calculation, where $h_{\text{BS}}$ and $h_{\text{MS}}$ are the \gls{bs} and \gls{mt} heights, respectively, $f_c$ is the carrier frequency, $c$ is the speed of light. On the other hand, NR's model includes an additional log-normal shadowing term. All details and formulas for the power loss computation are reported in Tab.~\ref{tab:path_loss}.

\subsection{Sionna~RT} 
Sionna~RT (v0.19.1) is a recently introduced differentiable ray tracing engine integrated into the Sionna link-level simulation library \cite{sionna}. It enables precise ray-based simulations, accounting for key propagation interactions such as reflection, diffraction, and diffuse scattering within the environment. It employs techniques from differentiable rendering and provides a versatile framework that supports the integration of communication channels into end-to-end optimization processes \cite{zhu2024toward}. 
Sionna~RT offers two ray-based simulation methods. \textit{i)} Exhaustive method: evaluates all possible combinations of 3D primitives and paths. While highly comprehensive, it becomes computationally prohibitive for scenarios involving high path depths (i.e. multiple interaction points per path) or a large number of surfaces in the scene; \textit{ii)} Shooting and bouncing: rays are launched into the scene by sampling a Fibonnacci lattice on the sphere. This technique efficiently computes propagation paths even in large scenes. It incorporates a sampling mechanism for ray launching, allowing users to balance simulation accuracy and computational efficiency.

\begin{figure}[!t]
\centerline{\includegraphics[width=0.9\linewidth]{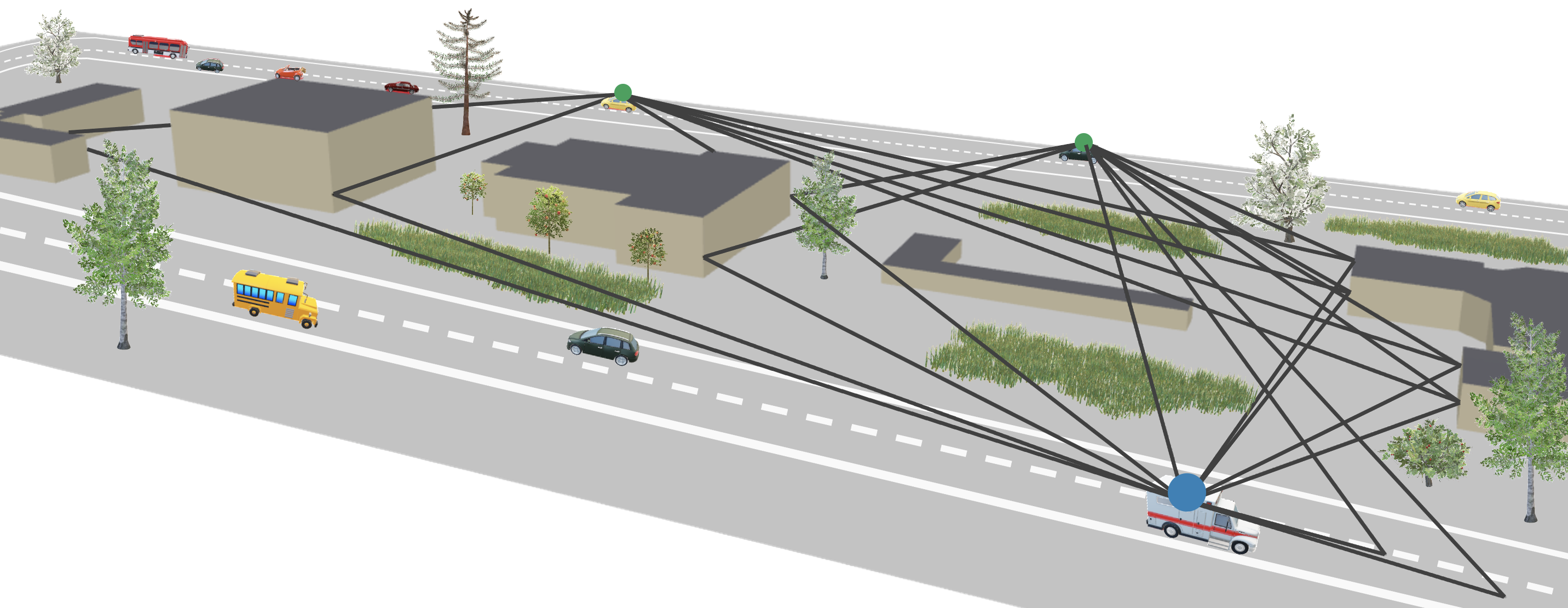}}
\caption{The considered 3D vehicular scenario, where moving vehicles are randomly connected or not. For testing purposes, every vehicle was supposed to be \gls{cav}.}
\label{scenario}
\end{figure}

\section{Impact of Ray Tracing on the Accuracy of a Full-Stack Simulation}
\label{sec:results}
\gls{v2v} scenarios are excellent test cases for ray tracing in simulations, as they effectively evaluate the improvements of the proposed solution under dynamic and complex conditions prone to \gls{nlos} due to vehicles and buildings. 
For this purpose, a \gls{v2v} simulation is carried out using ms-van3t~\cite{raviglione2024ms}\footnote{The source code of a ray tracing-enabled ms-van3t is available here:\textit{ https://github.com/robpegurri/ms-van3t-rt }}, an ns-3 native framework for standard-compliant vehicular communications which supports multiple \glspl{rat}, such as 802.11p, LTE \gls{v2v}, and NR \gls{v2v}.

A 90-second simulation of 20 \glspl{cav} exchanging \glspl{cam} with a rate ranging from 1~Hz to 10~Hz is performed. The physical scene represents a typical urban mobility scenario, as shown in Fig.~\ref{scenario}.

\subsection{Multi-\gls{rat} simulation}
\newcommand{\imwdt}{0.39}

\begin{figure}[!t]
    \centering
    \begin{subfigure}{\imwdt\textwidth} 
        \centering
        \includegraphics[width=\linewidth]{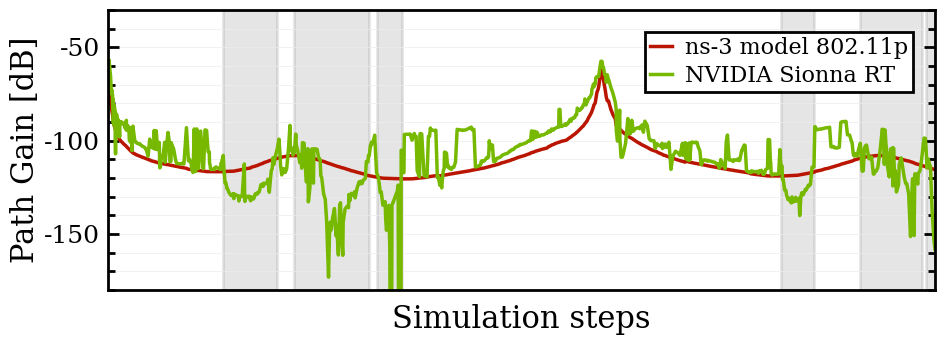}
        \caption{802.11p}
        \label{80211p}
        \vspace{-0.3cm}
    \hfill 
    \end{subfigure}
    \vspace{-0.3cm}
    \begin{subfigure}{\imwdt\textwidth}
        \centering
        \includegraphics[width=\linewidth]{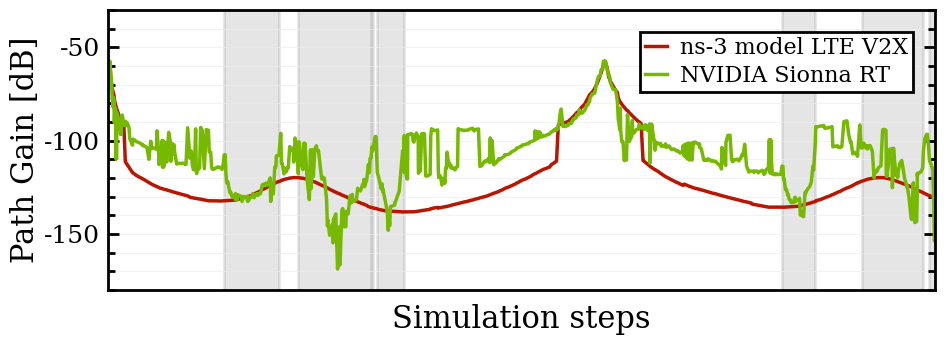}
        \caption{LTE \gls{v2v}}
        \label{LTE V2V}
    \hfill 
    \end{subfigure}
    \begin{subfigure}{\imwdt\textwidth}
        \centering
        \includegraphics[width=\linewidth]{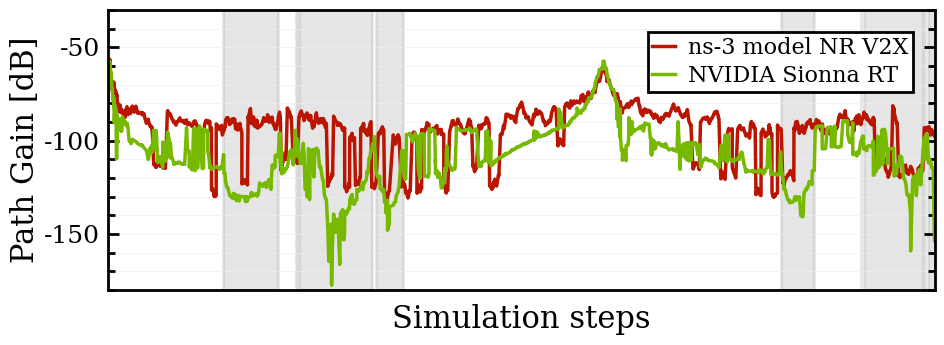}
        \caption{NR \gls{v2v}}
        \label{NR V2V}
    \hfill 
    \end{subfigure}
    
    \caption{ Computed path gain between two entities using three \glspl{rat} at the same carrier at 5.89~GHz. The highlighted sections of the graphs represent \gls{nlos} measurements.}
    \label{fig:powers_multiRAT}
\end{figure}
\begin{figure}[!t]
    \centering
    \begin{subfigure}{0.15\textwidth} 
        \centering
        \includegraphics[width=\linewidth]{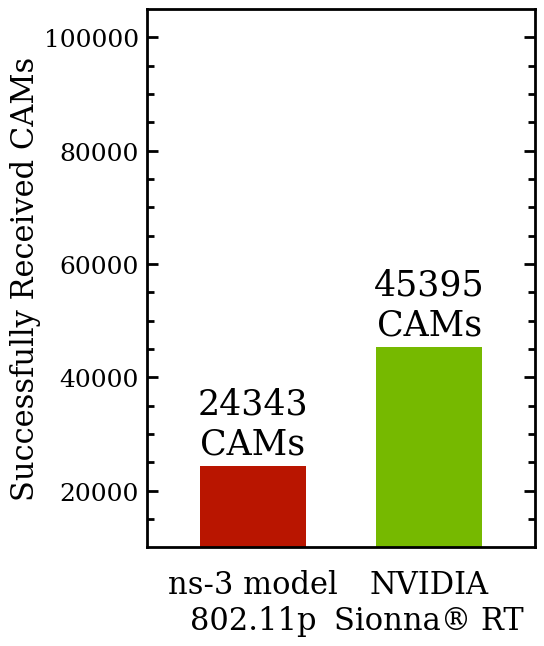}
        \caption{802.11p}
        \label{80211p_cam}
    \end{subfigure}
    \hfill 
    \begin{subfigure}{0.15\textwidth}
        \centering
        \includegraphics[width=\linewidth]{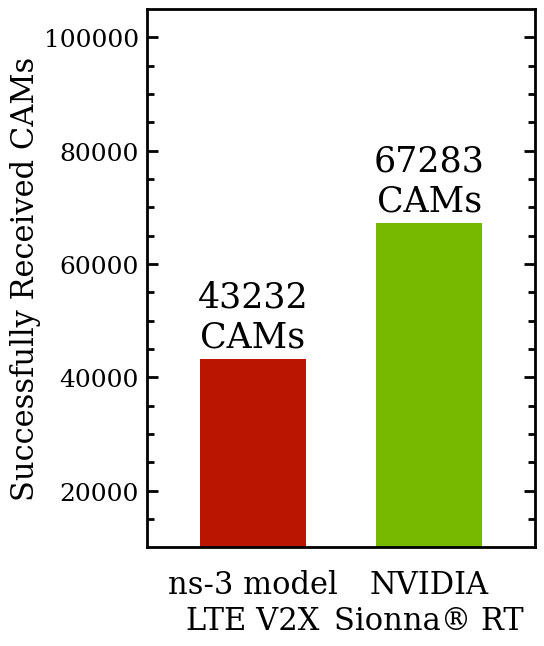}
        \caption{LTE \gls{v2v}}
        \label{LTE_cam}
    \end{subfigure}
    \hfill 
    \begin{subfigure}{0.15\textwidth}
        \centering
        \includegraphics[width=\linewidth]{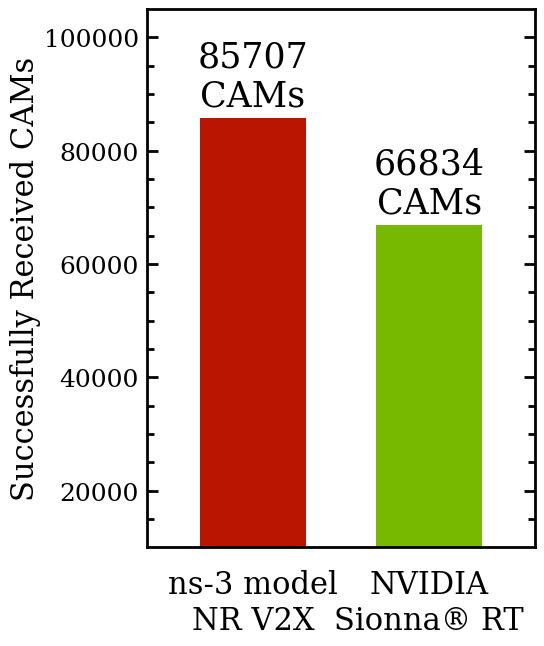}
        \caption{NR \gls{v2v}}
        \label{NR_cam}
    \end{subfigure}
    
    \caption{ Impact of ray tracing on successfully received \glspl{cam} across different \gls{rat} at 5.89~GHz.}
    \label{fig:multirat-comparison}
\end{figure}

\begin{figure} [!t]
    \centering
    \includegraphics[width=0.8\linewidth]{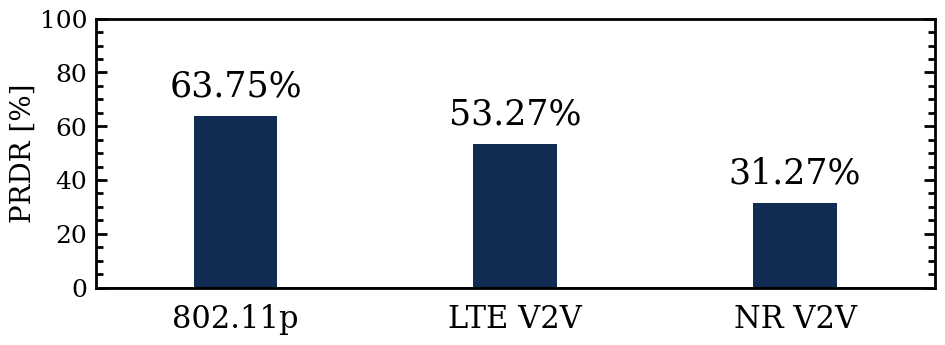}
    \caption{ Packet Reception Disagreement Ratio (PRDR), percentage of \glspl{cam} messages marked as delivered by ns-3 stochastic models incoherently with the ray tracer channel.} 
    \label{multirat-coherence}
\end{figure}

Figure~\ref{fig:powers_multiRAT} shows the computed path gains obtained by both ns-3 (based on Tab.~\ref{tab:channel_models} and \ref{tab:path_loss}) and Sionna~RT for a pair of moving vehicles at 5.89~GHz. The results are provided for 802.11p (Fig. \ref{80211p}), LTE \gls{v2v} (Fig. \ref{LTE V2V}), and NR \gls{v2v} (Fig. \ref{NR V2V}).

In LTE and NR, the probabilistic approach in  \mbox{ns-3} simulates fluctuations between \gls{los} and \gls{nlos} conditions, caused by obstructions such as buildings. On the other hand, the 802.11p model supposes the communication is always in \gls{los}. This results in a notable mismatch with ray tracing outputs.

The 802.11p model (Fig. \ref{80211p}) fails to account for \gls{nlos} conditions, leading to overestimated path gain peaks and pessimistic representation of the link performance. Conversely, the LTE \gls{v2v} model (Fig. \ref{LTE V2V}) wrongly estimates link conditions and gain values, treating the link as being in \gls{nlos} for nearly the entire simulation duration. The NR \gls{v2v} model (Fig. \ref{NR V2V}) is the most sophisticated among the three, producing gain values that align closely with ray tracing outputs for most of the simulation. However, it occasionally predicts false \gls{los} or \gls{nlos} conditions, introducing inaccuracies in the form of missing or unexpected dips in the gain values, resulting in an overall underestimation of the channel.

Figure~\ref{fig:multirat-comparison} demonstrates the impact of the aforementioned discrepancies at the network-level by considering the received packet counts. As expected, stock ns-3 channels for 802.11p and LTE \gls{v2v} lead to a lower estimated number of successfully received packets, while the stock ns-3 channel in NR \gls{v2v} causes the opposite. In LTE and NR, an overall higher number of delivered packets is observed due to their more robust error correction and recovery mechanisms.

To better highlight this aspect, we have quantified the level of disagreement between stock ns-3 and RT-enhanced ns-3 by means of the \gls{prdr}.
Let $\mathcal{S}$ and $\mathcal{R}$ the set of packets labeled as successfully received in stock ns-3 and RT-enhanced ns-3, respectively. Then $\mathcal{S}\Delta\mathcal{R}$ represents the set of packets received by either of the two simulators, but not both, namely the packets on which there is a disagreement. Then \gls{prdr} can be expressed as follows:
\begin{equation}
    PRDR = \left(\mathcal{S}\Delta\mathcal{R}\right)/\left(\mathcal{S}\cup \mathcal{R}\right)
\end{equation}

The \gls{prdr} reported in Fig.~\ref{multirat-coherence} highlights the tendency of stochastic models to misestimate network reliability. 
In particular, the stock ns-3 model for 802.11p inaccurately predicts 63.75\% of successfully delivered packets. This disagreement stems from channel overestimation and the complete disregard for the possibility of \gls{nlos} conditions. On the other hand, a lower disagreement of 53.27\% is still observed for the stock LTE model, now primarily due to its overestimation of the attenuation experienced in \gls{nlos} conditions. A slightly better agreement is achieved by the stock NR model with an incoherence percentage of 31.27\%, lower but still too far from the precision needed for \gls{dnt} applications.


\subsection{Multi-band simulation}

\begin{figure}[!t]
    \centering
    \begin{subfigure}{\imwdt\textwidth} 
        \centering
        \includegraphics[width=\linewidth]{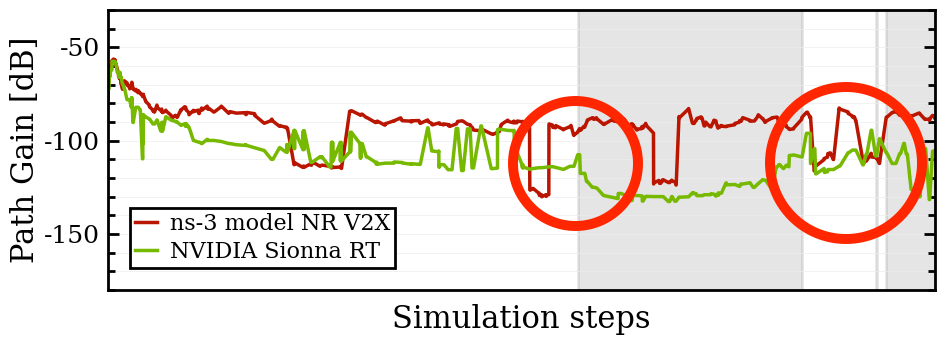}
        \caption{5.89~GHz (FR1)}
        \label{FR1_pw}
    \hfill 
    \end{subfigure}
    \begin{subfigure}{\imwdt\textwidth}
        \centering
        \includegraphics[width=\linewidth]{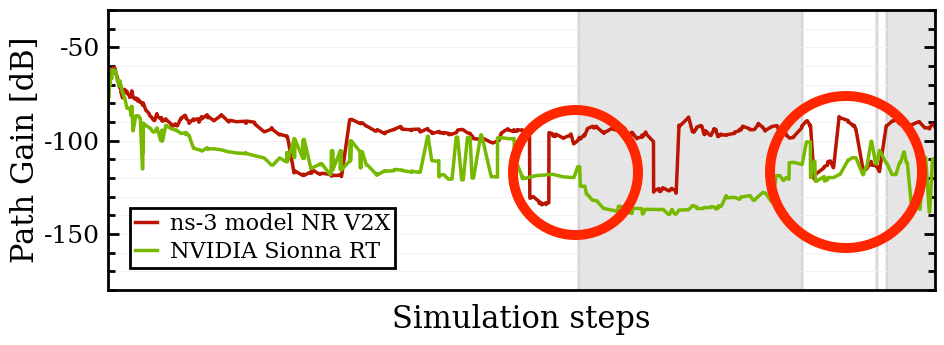}
        \caption{10~GHz (FR3)}
        \label{FR3_pw}
    \hfill 
    \end{subfigure}
    \begin{subfigure}{\imwdt\textwidth}
        \centering
        \includegraphics[width=\linewidth]{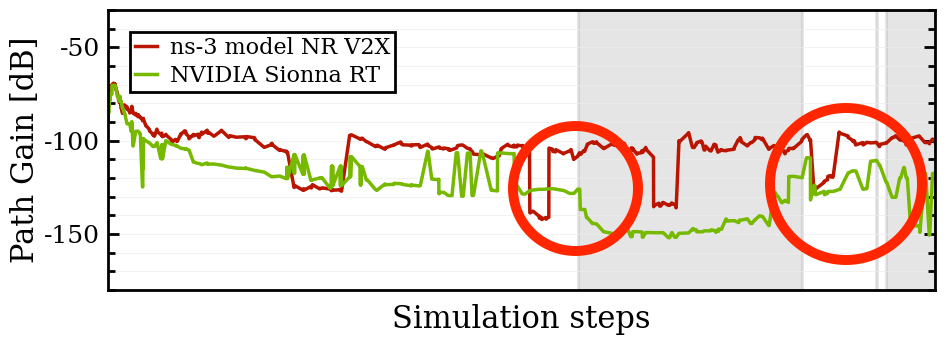}
        \caption{26~GHz (FR2)}
        \label{FR2_pw}
    \hfill 
    \end{subfigure}
    
    \caption{Path gains between two entities using NR \gls{v2v} across different frequencies. The highlighted sections of the graphs represent \gls{nlos} measurements.}
    \label{fig:powers_multiband}
\end{figure}

Figure~\ref{fig:powers_multiband} compares the path gains of the stock NR model with ray tracer results across three frequency bands: FR1 at 5.98~GHz (Fig.\ref{FR1_pw}), FR3 at 10~GHz (Fig.\ref{FR3_pw}), and FR2 at 26~GHz (Fig.~\ref{FR2_pw}). 
As mentioned previously, the stock NR \gls{v2v} model shows the lowest disagreement. However, \gls{los} or \gls{nlos} conditions are still probabilistically classified, introducing inaccuracies appearing as missing or unexpected dips in the gain values. 

On the other hand, the ray tracer incorporates coherent combination of paths. This method provides a more accurate representation of the channel, especially at higher frequencies. Consequently, the carrier frequency influences not only as a scaling factor but also by shaping the path gain trend, as evident from the differing peaks highlighted in red in Fig.~\ref{fig:powers_multiband}.

\begin{figure} [!t]
    \centering
    \includegraphics[width=0.8\linewidth]{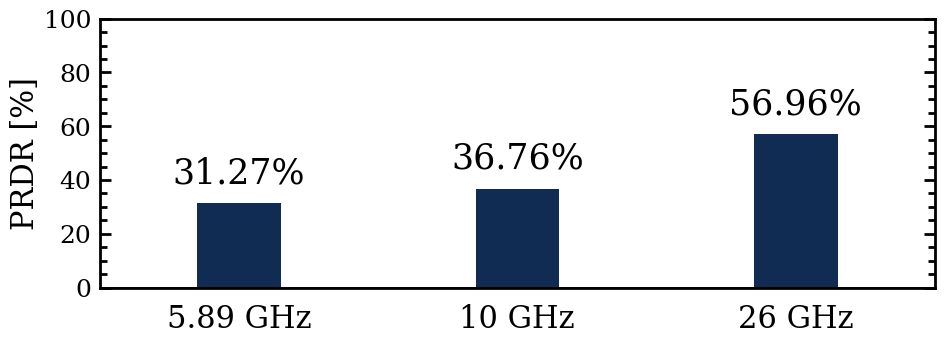}
    \caption{ Packet Reception Disagreement Ratio (PRDR), percentage of \glspl{cam} messages marked as delivered by ns-3 NR \gls{v2v} stochastic model incoherently with the ray tracer channel.} 
    \label{multiband-incoherence}
\end{figure}

The \gls{prdr} for these multi-band simulations is reported in Fig.~\ref{multiband-incoherence}, which reveals a discrepancy of up to 56\% at the FR2 band.  This substantial divergence, which increases with frequency, is attributed to the escalating influence of physical channel effects and propagation phenomena, primarily scattering and diffraction. These effects cannot be accurately modeled using stochastic approaches, resulting in a widening discrepancy between the simulations’ outcomes, particularly at higher frequency ranges such as FR2. Given the critical role of \gls{dnt} in high-frequency networks---where increased complexity demands greater accuracy--relying on RT-based simulations becomes essential for realistic and reliable modeling.

\section{Research Directions} \label{sec:directions}
This work demonstrates the potential and feasibility of integrating ray tracers into network simulations within the framework of \glspl{dnt}. While the results are promising, several challenges remain to be addressed to fully exploit this integration and advance the broader field. Below, we outline key research directions:

\textbf{Computation, Architecture, and Flexibility}: The computational demands of integrating ray tracers into network simulations are significant. Future research should focus on developing optimized architectures and ensuring flexibility to enable fully disaggregated and interoperable \gls{dnt} components at scale. Open interfaces will play a critical role in fostering interoperability. Additionally, improving computational efficiency is essential to reduce simulation runtimes, enabling \gls{dnt} systems to approach real-time execution speeds.

\textbf{Enhanced Utilization of Ray Tracing Information}: In the current integration, a stochastic path loss channel has been replaced with one generated by RT. While this improves accuracy, it does not make use of valuable information that RTs can provide, such as radio map data (e.g., angles of arrival, interference patterns, Doppler effects, and so on). Future work should aim to render such detailed information accessible to simulators, particularly for applications like coexistence studies where these details are crucial. At the same time, the network simulators themselves should be modified to effectively make use of such information. 

\textbf{Hardware Acceleration}: As this integration matures, we anticipate a shift toward offloading more computationally intensive functions---especially those in the PHY layer---to hardware accelerators such as GPUs. This evolution is necessary because general-purpose CPUs struggle with the high computational requirements of numerous PHY layer functions. Currently, these functions are either included in traditional simulators (slowing down simulations significantly) or abstracted (which limits their utility for certain applications). Hardware acceleration may strike a balance by maintaining simulation fidelity while improving performance. For instance, Sionna implements many PHY layer functions, such as waveform generation, MIMO and forward error corrections. These computationally intensive functions may thus be offloaded in future iterations of the proposed integration. 

\section{Conclusion} \label{sec:conclusions}

This work integrates ns-3 with Sionna~RT, creating the first open-source, full-stack, multi-RAT framework for 6G simulation. By using deterministic ray tracing, it overcomes stochastic model limitations, achieving higher accuracy in dynamic, multi-layer network analysis.  
Validated in a complex vehicular scenario, the framework revealed up to 65\% differences in application-layer performance, highlighting the critical role of precise channel modeling. Ray tracing consistently outperforms stochastic models under NLOS conditions, proving essential for realistic and reliable 6G network simulations.

\footnotesize
\section*{Acknowledgment}
This work was partially supported by the European Union - Next Generation EU under the Italian National Recovery and Resilience Plan (NRRP), Mission 4, Component 2, Investment 1.3,CUP D43C22003080001, partnership on “Telecommunications of the Future” (PE00000001 - program “RESTART”)

\bibliographystyle{IEEEtran}
\bibliography{Bibliography}

\end{document}